\documentclass[aps,twocolumn,showpacs,floatfix]{revtex4}
\usepackage{graphics}
\usepackage{epsfig}
\usepackage{subfigure}
\usepackage{amsmath,amsfonts,amssymb,graphicx}

\begin{document}
\title{Interplay between evolutionary game and network structure:\\
the coevolution of social net, cooperation and wealth
distribution}
\author{Jie Ren$^{1}$}
\author{Xiang Wu$^{1}$}
\author{Wen-Xu Wang$^{1,}$$^{2}$}
\author{Guanrong Chen$^{2}$}\email{gchen@ee.cityu.edu.hk}
\author{Bing-Hong Wang$^{1}$}
\affiliation{$^{1}$Department of Modern Physics, University of
Science and Technology of China, Hefei 230026, China\\
$^{2}$Department of Electronic Engineering, City University of
Hong Kong, Hong Kong SAR, China}
\date{\today}

\begin{abstract}
We study the interplay between evolutionary game and network
structure and show how the dynamics of the game affect the growth
pattern of the network and how the evolution of the network
influence the cooperative behavior in the game. Simulation results
show that the payoff-based preferential attachment mechanism leads
to the emergence of a scale-free structural property, $P(k)\sim
k^{-\gamma}$. Moreover, we investigate the average path length and
the assortative mixing features. The obtained results indicate
that the network has small-world and positive assortative
behaviors, which are consistent with the observations of some real
social networks. In parallel, we found that the evolution of the
underlying network structure effectively promotes the cooperation
level of the game. We also investigate the wealth distribution
obtained by our model, which is consistent with the Pareto law in
the real observation. In addition, the analysis of the generated
scale-free network structure is provided for better understanding
the evolutionary dynamics of our model.
\end{abstract}

 \maketitle
\section{Introduction}

Game theory provides a useful framework for describing the
evolution of systems consisting of selfish individuals
\cite{game1,game2,game3}. The prisoner's dilemma game (PDG) as a
metaphor for investigating the evolution of cooperation has drawn
considerable attention \cite{PD1,PD2}. In the PDG, two players
simultaneously choose whether to cooperate or defect. Mutual
cooperation results in payoff $R$ for both players, whereas mutual
defection leads to payoff $P$ gained both. If one cooperates while
the other defects, the defector gains the highest payoff $T$,
while the cooperator bears a cost $S$. This thus gives a simply
rank of four payoff values: $T>R>P>S$. One can see that in the
PDG, it is best to defect regardless of the co-player's decision
to gain the highest payoff $T$. However, besides the widely
observed selfish behavior, many natural species and human being
show the altruism that individuals bear cost to benefit others.
These observation brings difficulties in evaluating the fitness
payoffs for different behavioral patterns, even challenge the rank
of payoffs in the PDG. Since it is not suitable to consider the
PDG as the sole model to discuss cooperative behavior, the
snowdrift game (SG) has been proposed as possible alternative to
the PDG, as pointed out in Ref \cite{Hauert}. The main difference
between the PDG and the SG is in the order of $P$ and $S$, as
$T>R>S>P$ in the SG. This game, equivalent to the hawk-dove game,
is also of much biological interest \cite{SG1,SG2}. However, the
original PDG and SG cannot satisfyingly reproduce the widely
observed cooperative behavior in nature and society. This thus
motivates numerous extensions of the original model to better
mimic the evolution of cooperation in the real world
\cite{Nowak2,Nowak3,Nowak4,Lieberman}.

Since the spatial structure is introduced into the evolutionary
games by Nowak and May \cite{Nowak1}, there has been a continuous
effort on exploring effects of spatial structures on the
cooperation \cite{Doebeli1,Doebeli2,Hauert}. It has been found
that the spatial structure promotes evolution of cooperation in
the PDG \cite{Nowak1}, while in contrast often inhibits
cooperative behavior in the SG \cite{Hauert}. In recent years,
extensive studies indicate that many real networks are far
different from regular lattices, instead, show small-world and
scale-free topological properties. Hence, it is naturally to
consider evolutionary games on networks with these kinds of
properties \cite{Abramson,Kim,Masuda,WZX,Kim2,DoubleZheng,Jie}. An
interesting result found by Santos and Pacheco is that
``Scale-free networks provide a unifying framework for the
emergence of cooperation" \cite{Santos}. So far, most studies of
evolutionary games over networks are based on static network
structure. However, it has been pointed out that the network
structure may coevolve with the game
\cite{Interplay1,Interplay11,Interplay2,Interplay3,Interplay4},
where each individual would choose its co-players to gain more
benefits, inducing the evolution of their relationship network.
Some previous works about weighted networks suggest that it is
indeed the traffic increment spurs the evolution of the network to
maintain the system's normal and efficient functioning
\cite{traffic1,traffic2}. From this perspective, in the present
paper we propose an evolutionary model with respect to the
interplay between the evolutions of the game and the network for
characterizing the dynamics of some social and economic systems.

In our model, the SG is adopted for its more general
representation of the realism and evolutionary cooperative
behavior. Since growth is a common feature among networked systems
\cite{BAreview}, we assume that the network continuously grows by
adding new agents to the existent network on the basis of the
payoff preferential attachment. We focus on the evolution of the
network structure together with the emergence and persistence of
cooperation. Simulation results show that the obtained networks
follow a power-law distribution, $p(k)\sim k^{-\gamma}$, with
exponent $\gamma$ tuned by a model parameter. The average distance
of the network scales logarithmically with the network size, which
indicates the network has a small-world effect. Interestingly, the
assortative mixing properties generated by our model demonstrate
that the model can well mimic social networks. In parallel, with
the extension of the network, the density of cooperators increases
and approaches a stable value, which gives a new explanation for
the emergence and persistence of cooperation. We also explore the
wealth distribution, where the so-called wealth is the accumulated
payoff distribution of each individual. The Pareto law is well
reproduced by our model. At last, we provide analyses for the
obtained scale-free network structures.

The paper is arranged as follows. In the following section, we
describe the model in detail, in Sec. III, simulation results and
correspondent analytical ones are provided, and in Sec. IV, the
work is concluded.

\section{The model}
Let us introduce briefly the SG first. Consider two drivers are
trapped in two side of a snowdrift. Each driver has two possible
selections, either shoving the snowdrift (cooperator-C) or
remaining in the car and do nothing (defect-D). If both cooperate,
they could be back home on time, so that each will gain a reward
of $b$, whereas mutual defection results in still blocked by the
snowdrift and each gets a payoff $P=0$. If only one driver shovels
(takes C), then both drivers can be back home. The driver taking D
gets home with do nothing and hence gets a payoff $T=b$, while the
driver taking C gains a ``sucker" payoff of $S=b-c$. Thus, the
rank of four payoff values is $T>R>S>P$. Following common
practice, the SG is rescaled with $R=1$, $T=1+r$ and $S=1-r$,
where $r$ is a tunable parameter ranging from $0$ to $1$. Hence,
the payoffs can be characterized by a single parameter for
convenient study.

\begin{figure}
\scalebox{0.80}[0.80]{\includegraphics{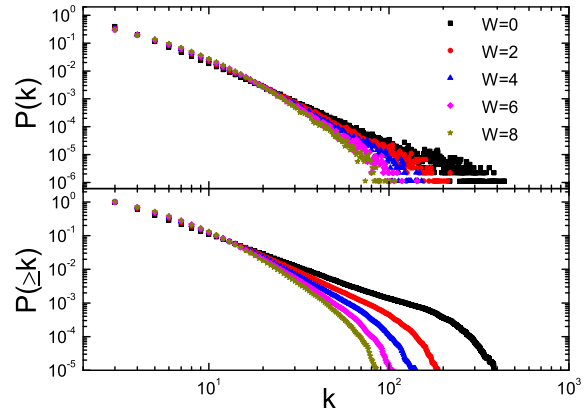}}
\caption{\label{fig:epsart} (color online). Degree distributions
and relevant cumulative degree distributions for different values
of $W$ are shown in the top panel and the bottom, respectively.
The network size $N$ is $10000$. Each distribution is obtained by
averaging over $10$ distinct simulation. The degree distribution
is nearly independent of $r$.}
\end{figure}

Our model starts from $m_0$ nodes randomly connected with
probability $p$, each of which represents a player (In the
following, we fix $m_0=10, p=0.6$ and examine that it has no
influence on our results in present work). Initially, the nodes
are randomly assigned to be either strategy C or D with $50-50$
percentages. Players interact with all their neighbors
simultaneously and get payoffs according to the preset payoff
parameter. The total payoff of a certain player is the sum over
all its encounters. Then, every node $i$ randomly selects a
neighbor $j$ at the same time for possible updating its strategy.
The probability that $i$ follows the strategy of the selected node
$j$ is determined by the total payoff difference between them,
i.e.,
\begin{equation}
p_{ij}=\frac{1}{1+\exp[(M_i-M_j)/T]},
\end{equation}
where $M_i$ and $M_j$ are the total payoffs of $i$ and $j$ at the
moment of the encounter. Here, $T$ characterizes ``noise",
including bounded rationality, individual trials, errors in
decision, \emph{etc}. It should be noted that $T$ here plays a
different role comparing with the cases of adopting the normalized
payoff difference. In parallel, $T$ does not play the same role
for different network sizes \cite{T1,T2,Szabo4,Szabo5}. Since in
our model, the network size gradually grows, it is not easy to
keep the effect of $T$ unchanged at every time step. For
simplicity, we fix $T$ to $0.1$ during the evolution of the
network.

Here, we adopt the synchronous updating rule. After each step that
players update their strategies, a new individual is added into
the network with $m\leq m_0$ (we fix $m=3$ for convenience) links
preferentially attached to existent nodes of higher payoffs, i.e.,
\begin{equation}
\Pi_{new \rightarrow i}=\frac{M_i+W}{\sum_j (M_j+W)},
\end{equation}
where $M_i$ and $M_j$ are the total payoffs of $i$ and $j$
obtained in the interaction process. $W$ is a tunable parameter,
which reflects the original payoff values of players when they
join into the game system. For simplicity, we set $W$ be a
constant. The payoff-based preferential selection takes into
account the ``rich gets richer" characteristic and couples the
dynamics of the evolutionary game and the evolution of the
underlying network. After a new player joins into the network, the
new one randomly choose strategy C or D and all old players
preserve their strategies for the game in the next round. Then,
repeat the above procedures, and the network size gradually grows.

\section{Simulation and analytical results}
\subsection{The evolution of networks}

\begin{figure}
\scalebox{0.80}[0.80]{\includegraphics{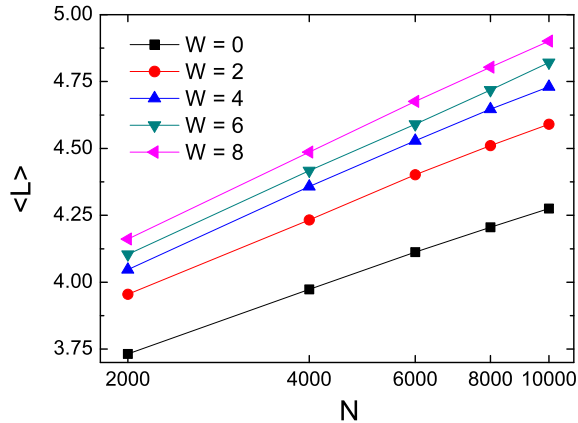}}
\caption{\label{fig:epsart} (color online). Average distance
$\langle L \rangle$ as a function of network size $N$. Each data
point is obtained by averaging over $10$ network realizations.
These results are independent of parameter $r$.}
\end{figure}

Numerical simulations are performed to quantify the structural
properties of the obtained networks. In Figure 1, we show the
degree distribution and correspondent cumulative degree
distribution $P(\geq k)$, in networks of size $N=10000$. The
distributions clearly exhibit power-law behaviors, $P(k)\sim
k^{-\gamma}$, in a broad range of degrees with a fat tail for very
large degrees. Besides, for the cumulative degree distribution
$P(\geq k)$, a cut-off at very large degrees is observed for each
distribution, which corresponds to the fat tail range. The
cumulative degree distribution provides a clear picture of the
power-law behavior. These results indicate that the empirically
observed scale-free structure can be generated from the coupling
of the game and the evolution of the network, which may be an
explanation for the heterogenous structure of many social and
economical networked-systems. Moreover, the exponent $\gamma$ is a
function of $W$, which makes our model more general for mimicking
a variety of real networks. We have checked that the parameter $r$
has slight effect on $\gamma$, while $W$ plays a major role.
Analytical results of the power-law distribution will be given
after the discussion of the correlation between individuals's
payoffs and the degrees of nodes occupied by them.

\begin{figure}
\scalebox{0.80}[0.80]{\includegraphics{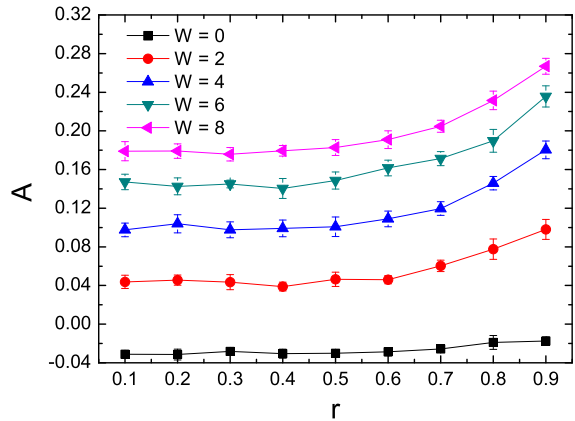}}
\caption{\label{fig:epsart} (color online). Assortative mixing
coefficient $A$ as a function of parameter $r$ for different
values of $W$. The network size $N$ is $10000$. Each data point is
obtained by averaging over $10$ network realizations.}
\end{figure}

Average path length is a key measure for quantifying the
small-world effect, which is widely observed in the real world
\cite{Newman}. The average path length $\langle L \rangle$ of a
network is defined as
\begin{equation}
\langle L \rangle=\frac{2}{N(N+1)} \sum_{i\geq j} d_{ij},
\end{equation}
where $d_{ij}$ is the shortest path length from node $i$ to node
$j$, and $N$ is the network size. We perform simulations on
$\langle L \rangle$ as a function of network size for different
values of parameter $W$. Each data point is obtained by averaging
over $10$ different network realizations. Figure 2 shows that for
all the values of $W$, $\langle L \rangle$ increases
logarithmically with the growth of the network, but the slopes for
different $W$ show slight differences. These results demonstrate
that the small-world effect can be reproduced by the proposed
model.

Another important structural feature useful for measuring the
correlation among nodes of a network is the assortative mixing
coefficient $A$, or called degree-degree correlation
\cite{mixing1,mixing2}, which is defined as follows:
\begin{equation}
A=\frac{M^{-1}\sum_i j_ik_i-[M^{-1}\sum_i
\frac{1}{2}(j_i+k_i)]^2}{M^{-1} \sum_i \frac{1}{2}(j_i^2+k_i^2)
-[M^{-1}\sum_i \frac{1}{2}(j_i+k_i)]^2},
\end{equation}
where $j_i$ and $k_i$ are the degrees of the two nodes at the end
of the $i$th edge, with $i=1,\cdots, M$ ($M$ is the total number
of edges of the observed graph). Two main classes of possible
correlations have been observed in the real world: assortative
behavior if $A>0$, which indicates that large-degree nodes are
preferentially connected with other large-degree nodes, and
disassortative if $A<0$, which denotes that links are more easily
built between large-degree nodes and small-degree ones. As
demonstrated in Ref \cite{mixing1}, almost all social networks
show positive values of $A$, while others, including technological
and biological networks, show negative $A$. However, the mechanism
that leads to the basic difference between these two classes
networks remains unclear \cite{why}. We calculate the assortative
mixing coefficient $A$ to check whether the generated networks by
our model are suitable representations of social systems. Figure 3
shows $A$ as a function of parameter $r$ for different values of
$W$. One can see that assortative behavior occurs as $W$
increases. For each value of $W$, $A$ shows slight changes in the
cases of low values of $r$, while for large $r$, assortative
mixing is enhanced with the same value of $W$. The reported
results demonstrate that the networks generated by our model can
well capture the key distinguished structural property of social
networks.

\begin{figure}
\scalebox{0.80}[0.80]{\includegraphics{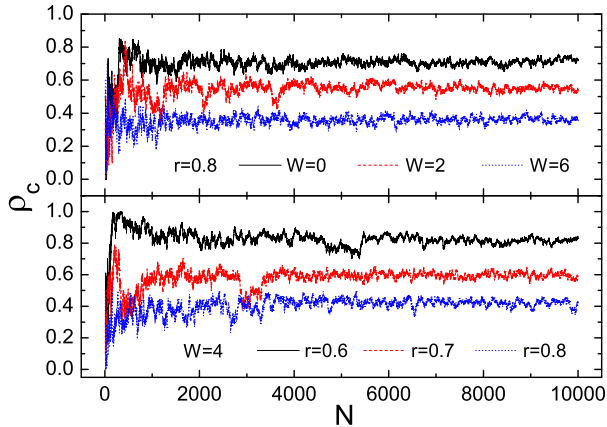}}
\caption{\label{fig:epsart} (color online). Cooperation density
$\rho_c$ as a function of network size $N$ for different $W$ with
fixing $r$ in the top panel and for different $r$ with fixing $W$
in the bottom panel.}
\end{figure}

\subsection{Cooperation and wealth distribution}
So far we have studied the evolution of the underlying network
structure influenced by the game. Next, we turn to the effect of
changing network structure on the cooperative behavior in the
game. The key quantity for measuring the cooperation level of the
game is the density of cooperators, $\rho_c$ \cite{Kim}. Figure 4
shows the time series of $\rho_c$ for different values of $W$ and
$r$. One can find that after a short period of temporary behavior,
$\rho_c$ reaches a stable value with small fluctuations around the
average value. Hence, $\rho_c$ for each pair of $W$ and $r$ can be
calculated by averaging a period time after the system enter a
steady state. Here, the network size $N$ is equal to the
evolutionary time step, since each time an individual joins into
the system. In Fig. 5, we report the $\rho_c$ depending on $r$ for
different values of $W$. Each data point is obtained by averaging
over 10 different simulations with an average from $N=5000$ to
$10000$ for each simulation. One can see in Fig. 5, in the case of
$r\leq 0.4$, $\rho_c$ shows no difference for distinct $r$. While
for $r>0.4$, the lower the value of $r$, the higher the
cooperation level. It has been known that scale-free networks
favor the emergence and persistence of cooperation \cite{Santos}.
Thus, the fact that $\rho_c$ in our model is larger than that of
well-mixed cases is attributed to the emergence of scale-free
structural properties. At the very beginning, the network evolves
from a core of random-like structure, in which cooperation cannot
dominate in the game. While as the network gradually grows,
power-law degree distribution emerges, which leads to a sharp
increase of $\rho_c$, as shown in Fig. 4 from $N=0$ to $500$. The
inhibited cooperation by the increment of $W$ is also ascribed to
the weakened heterogeneity of degree distribution. As displayed in
Fig. 1, lower value of $W$ corresponds to stronger heterogeneity
of degree distribution reflected by the longer fat tail. The above
discussion gives a thorough picture that it is the growth and the
payoff-based preferential attachment that produce scale-free
network structures and meanwhile, the generated heterogeneity of
degree distribution effectively promotes the emergence and
persistence of cooperation.

\begin{figure}
\scalebox{0.80}[0.80]{\includegraphics{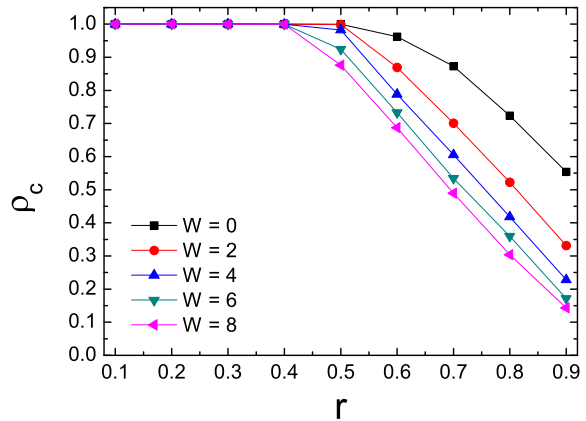}}
\caption{\label{fig:epsart} (color online). Cooperator density
$\rho_c$ as a function of payoff parameter $r$, for different $W$.
Each data point is obtained by averaging over 10 different
simulations with an average from $N=5000$ to $10000$ for each
simulation.}
\end{figure}

Generally speaking, cooperation and defection are prototypical
actions in economical systems. Hence, evolutionary games may be
suitable paradigms for studying and characterizing the phenomena
observed in economical systems with players represented by agents.
In such systems, a well-known and extensively studied phenomenon
is the wealth distribution of agents which follows the Pareto law
in the high-income group \cite{pareto,US,JP}. In order to check
the validity of our model for understanding economical behavior,
we investigate the wealth distribution by adopting the present
evolutionary model, where the wealth of an agent is naturally
represented by the accumulated payoff over time steps. Figure 6
reports the accumulated distribution of accumulated payoff $P_c$
in the whole population for different model parameters $r$ and
$W$. One can see that power-law distribution can be observed in a
wide range of $P_c$, while the wealth distributions behave as
exponential corrections in the zone of low values of $P_c$, which
is in accordance with the empirical evidence. Moreover, in the
left panel of Fig. 6, $W$ has strong influence on the exponent of
power-law distribution and higher value of $W$ corresponds to
larger exponent. In contrast, in the right panel, $r$ nearly has
no effect on the wealth distribution. The correlation between $W$
and the exponent of wealth distribution makes our model general
for reproducing the empirical observation.

\begin{figure}
\scalebox{0.80}[0.80]{\includegraphics{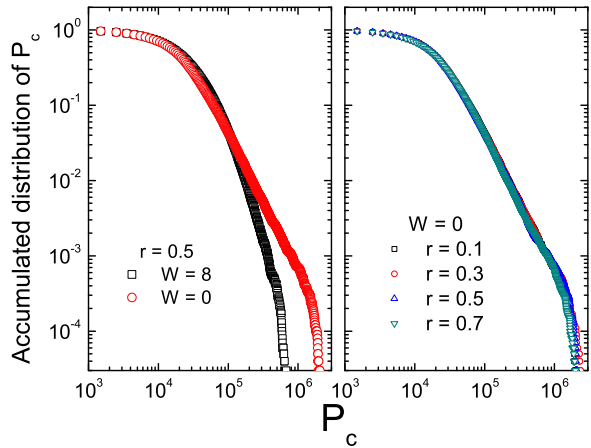}}
\caption{\label{fig:epsart} (color online). Accumulated wealth
distribution for different $W$ with fixing $r$ in the left panel
and for different $r$ with fixing $W$ in the right panel. The
so-called wealth $P_c$ is the accumulated payoff over times of
each individual. The results are obtained by averaging over $10$
network realizations when the network size $N$ reaches $10000$.
The exponents of the power-law distribution in the left panel for
$W=0$ and $W=8$ are $1.80$ and $2.89$, respectively. The empirical
data of USA and Japan are $1.6$ and $1.8-2.2$, respectively
\cite{US,JP}. Hence, the real observations can be reproduced by
our model by tuning the value of $W$.}
\end{figure}

\subsection{Analysis}
In the following, we provide some analysis for the scale-free
network structure induced by the payoff-preferential attachment
via considering the correlation between the accumulated payoff
$P_c$ of individuals and their correspondent degrees $k$. As shown
in Fig. 7, $P_c(k)$ is a good linear function of $k$ with slope
depending on $r$ in the simulations. Using the mean-field
approximation, a node with degree $k$ may have $k\rho_c$
cooperative neighbors and $k(1-\rho_c)$ defectors and itself may
be cooperator with probability $\rho_c$ and defector with
$1-\rho_c$. Thus, at time step $t$, its payoff can be calculated
as
\begin{eqnarray}
M_k(t)&=&\rho_c(t)\cdot k\rho_c(t)\cdot
R\nonumber\\&+&\rho_c(t)\cdot k(1-\rho_c(t))\cdot S \nonumber\\
&+&(1-\rho_c(t))\cdot k\rho_c(t)\cdot T \nonumber\\
&+&(1-\rho_c(t))\cdot k(1-\rho_c(t))\cdot P
\end{eqnarray}
By substituting the elements of payoff matrix of the SG in Eq.
(5), where $R=1$, $S=1-r$, $T=1+r$ and $P=0$, Eq. (5) is
simplified to
\begin{equation}
M_k(t)=k\rho_c(t)(2-\rho_c(t)).
\end{equation}
Then, we get the accumulated payoff
\begin{eqnarray}
P_c(k)&=&\sum_tM_k(t)\nonumber\\&=&k\sum_t\rho_c(t)(2-\rho_c(t)) \nonumber\\
&\propto& k\rho_c(2-\rho_c),
\end{eqnarray}
where the approximation is ascribed to the fact that $\rho_c(t)$
quickly reaches a stable value which is almost independent of $t$
(as shown in Fig. 4), so that $\rho_c(t)$ are replaced by $\rho_c$
for simplicity. The inset of Fig. 7 illustrates the comparison
between the simulation results and the analytical ones on the the
normalized slopes vs $r$. The theoretical predictions are
calculated by substituting the simulation results of $\rho_c$ (in
Fig. 5) into Eq. (7). The theoretical results are consistent with
simulations.

\begin{figure}
\scalebox{0.80}[0.80]{\includegraphics{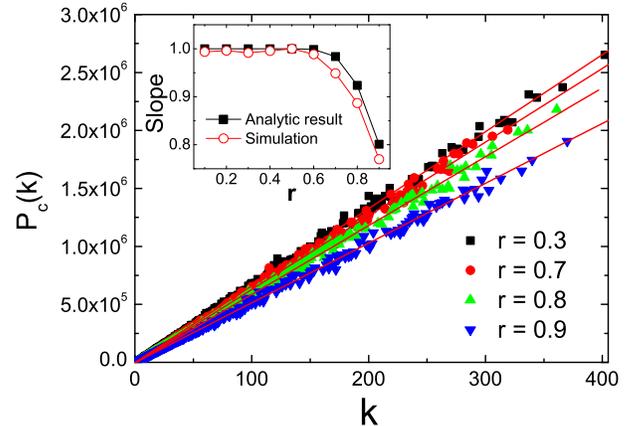}}
\caption{\label{fig:epsart} (color online). Correlation between
the accumulated payoff of each individual $P_c(k)$ and its degree
$k$ for different values of $r$. The results are obtained by
averaging over $10$ network realizations. $P_c(k)$ shows a linear
function of $k$. Simulation and analytical results of the
normalized slope of each line depending on $r$ are displayed in
the inset. The network size $N$ is $10000$.}
\end{figure}

Accordingly, considering the relation between the payoff $M_i$ of
an individual $i$ and its degree $k_i$ in Eq. (6), we give the
evolution equation of the degree of a given node
\begin{eqnarray}
\frac{dk_i}{dt}&=&\frac{m(M_i+W)}{\sum_{j=1}^t
(M_j+W)}\nonumber\\&=&\frac{m(k_i\rho_c(2-\rho_c)+W)}{2mt\rho_c(2-\rho_c)+Wt}\nonumber\\
&=&\frac{k_i+\frac{W}{m\rho_c(2-\rho_c)}}{(2+\frac{W}{m\rho_c(2-\rho_c)})t}.
\end{eqnarray}
Thus, we get $k_i(t)\sim t^\alpha$ with
$\alpha=1/(2+\frac{W}{m\rho_c(2-\rho_c)})$. Then, in the infinite
size, the degree probability distribution can be acquired by
$P(k)\sim\int_{0}^\infty\delta(k-k_i(t))dt\sim k^{-\gamma}$ with
\begin{equation}
\gamma=1+1/a=3+\frac{W}{m\rho_c(2-\rho_c)}.
\end{equation}
Though this expression is somewhat rough because of several
approximations, such as $\rho_c\approx\rho_c(t),
t\rightarrow\infty$, Eq. (9) can qualitatively describe the
power-law degree distribution of the generated network.

In addition, we should mention a network model proposed by
Dorogovtsev and Mendes \cite{IA}, which is related to the present
work. In such model, by introducing the ``initial attractiveness"
(IA) to each node, power-law degree distributions can be generated
together with the exponent of the distribution controlled by
strength of the IA. Very interestingly, the IA plays a significant
role in the emergence of the assortative mixing property
\cite{Assort}. In our model, the parameter $W$ may play the same
role as that of the IA in the perspective of assortative feature.
The introduction of $W$ enlarges the probability of poor players
being connected by the new one in the growth process. Moreover,
there is an approximately positive correlation between the payoff
and the degree of a given individual. Hence, $W$ enhances the
connecting probability between small-degree individuals that
results in the assortative mixing behavior. However, in our model,
the degree distribution is not only controlled by $W$ and $m$, but
also by $\rho_c$, as obtained in Eq. (9). Our model couples the
dynamical process of the SG and the evolution of the network,
which leads to the difference between our model and the network
model of Dorogovtsev and Mendes.

\section{conclusion}
In summary, we have studied the interplay of the evolutionary game
and the relevant network structure. Simulation results indicate
that both scale-free structural property and high cooperation
level result from the interplay between the game and the network.
Moreover, the resultant networks reproduce some typical features
of social networks, including small-world and positive assortative
mixing properties. The investigation of the wealth distribution of
players indicates the validity of our model in mimicking the
dynamical behavior of economical systems.

However, some issues still remain unclear and deserve further
study, such as the evolution of connections among existing nodes,
as discussed in previous works \cite{DM,MS1,why}. On the other
hand, in the present work, we only consider the case of ``birth"
of new players, which leads to the growth of the network. While,
in social systems, ``death" and ``aging" are also important events
and a previous work has already pointed out that the aging effect
plays a significant role in the evolution of network structures
\cite{aging}. Therefore, there is a need to consider the death and
aging processes for better characterizing the evolutionary
dynamics of social systems in the future study.


\begin{thebibliography}{ref1}
\bibitem{game1} A. M. Colman, {\sl Game Theory and its Applications in the Social and Biological
Sciences} (Butterworth-Heinemann, Oxford, 1995).

\bibitem{game2} J. Hofbauer and K. Sigmund,  {\sl Evolutionary Games and Population
Dynamics} (Cambridge University Press, Cambridge, U.K., 1998).

\bibitem{game3} R. Trivers, {\sl Social Evolution} (Cummings, Menlo Park,
1985).

\bibitem{PD1} R. Axelrod and W. D. Hamilton, Science \textbf{211}, 1390 (1981)

\bibitem{PD2} R. Axelrod, {\sl The Evolution of Cooperation} (Basic books, New York,
1984).

\bibitem{Hauert} C. Hauert and M. Doebeli, Nature \textbf{428}, 643
(2004).

\bibitem{SG1} R. Sugden, {\sl The Economics of Rights, Co-operation and
Welfare} (Blackwell, Oxford, U.K., 1986).

\bibitem{SG2} J. M. Smith, {\sl Evolution and the Theory of Games}
(Cambridge University Press, Cambridge, UK, 1982).



\bibitem{Nowak2} M. Nowak and K. Sigmund, Nature (London)
\textbf{355}, 250 (1992).

\bibitem{Nowak3} M. Nowak and K. Sigmund, Nature (London)
\textbf{364}, 1 (1993).

\bibitem{Nowak4} M. Nowak, A. Sasaki, C. Taylor, and D. Fudenberg,
Nature (London) \textbf{428}, 646 (2004).

\bibitem{Lieberman} E. Lieberman, C. Hauert, and M. Nowak, Nature
(London) \textbf{433}, 312 (2005).

\bibitem{Nowak1} M. Nowak and R. M. May, Nature (London) \textbf{359}, 826
(1992); Int. J. Bifurcation Chaos Appl. Sci. Eng. \textbf{3}, 35
(1993).

\bibitem{Doebeli1} M. Doebeli and N. Knowlton, Proc. Natl. Acad.
Sci. USA \textbf{95}, 8676 (1998).


\bibitem{Doebeli2} M. Doebeli, C. Hauert and T. Killingback, Science \textbf{306},
859 (2004).





\bibitem{Abramson} G. Abramson and M. Kuperman, Phys. Rev. E
\textbf{63}, 030901(R) (2001).

\bibitem{Kim} B. J. Kim, A. Trusina, P. Holme, P. Minnhagen, J. S.
Chung, and M. Y. Choi, Phys. Rev. E \textbf{66}, 021907 (2002).

\bibitem{Masuda} N. Masuda and K. Aihara, Phys. Lett. A
\textbf{313}, 55 (2003).

\bibitem{WZX} Z. X. Wu, X. J. Xu, Y. Chen, and Y. H. Wang, Phys.
Rev. E \textbf{71}, 037103 (2005).

\bibitem{Kim2} H. Hong, B. J. Kim, M. Y. Choi, and H. Park, Phys.
Rev. E \textbf{72}, 041906 (2005).

\bibitem{DoubleZheng} L.-X. Zhong, D.-F. Zheng, and B. Zheng, and
P. M. Hui, arXiv: physics/0602039.

\bibitem{Jie} J. Ren, W.-X. Wang, G. Yan, and B.-H. Wang, arXiv:
physics/0603007.

\bibitem{Santos} F. C. Santos and J. M. Pacheco, Phys. Rev. Lett.
\textbf{95}, 098104 (2005).



\bibitem{Interplay1} H. Ebel and S. Bornholdt, Phys. Rev. E \textbf{66},
056118 (2002).

\bibitem{Interplay11} M. G. Zimmermann, V. M. Egu\'{i}luz and M. S. Miguel,
Phys. Rev. E \textbf{69}, 065102 (2002).

\bibitem{Interplay2} M. G. Zimmermann and V. M. Egu\'{i}luz, Phys.
Rev. E \textbf{72}, 056118 (2005).

\bibitem{Interplay3} V. M. Egu\'{i}luz, M. G. Zimmermann, C. J.
Cela-Conde, and M. San Miguel, Am. J. Sociol. \textbf{110}, 977
(2005).

\bibitem{Interplay4} P. Holme and G. Ghoshal, Phys. Rev. Lett.
\textbf{96}, 098701 (2006).



\bibitem{traffic1} A. Barrat, M. Barth\'elemy, and A. Vespignani, Phys.
Rev. Lett. \textbf{92}, 228701 (2004).

\bibitem{traffic2} W.-X. Wang, B.-H. Wang, B. Hu, G. Yan, and Q. Ou,
Phys. Rev. Lett. \textbf{94}, 188702 (2005).



\bibitem{BAreview} R. Albert and A.-L. Barab\'asi, Rev. Mod. Phys. \textbf{74}, 47 (2002).


\bibitem{T1} A. Traulsen, J. C. Claussen and C. Hauert, Phys. Rev. Lett.
\textbf{95}, 238701 (2005).

\bibitem{T2} J. C. Claussen and A. Traulsen, Phys. Rev. E.
\textbf{71}, 025101 (2005).

\bibitem{Szabo4} G. Szab\'{o} and J. Vukov, Phys. Rev. E
\textbf{69}, 036107 (2004).

\bibitem{Szabo5} J. Vukov and G. Szab\'{o}, Phys. Rev. E
\textbf{71}, 036133 (2005).

\bibitem{Newman} M. E. J. Newman, SIAM Review \textbf{45}, 167 (2003).

\bibitem{mixing1} M. E. J. Newman, Phys. Rev. Lett. \textbf{89},
208701 (2002).

\bibitem{mixing2} M. E. J. Newman, Phys. Rev. E \textbf{67},
026126 (2003).

\bibitem{why} W.-X. Wang, B. Hu, B.-H. Wang, and G. Yan, Phys.
Rev. E \textbf{73}, 016133 (2006).



\bibitem{pareto} V. Pareto, {\sl Le Cours d'Economique Politique}
(Macmillan, Lausanne, Paris, 1987).

\bibitem{US} A. A. Dragulescu, V. M. Yakovenko, Physica A
\textbf{299}, 213 (2001).

\bibitem{JP} S. Moss de Oliveira, P. M. C. de Oliveira, D. Stauer,
{\sl Evolution, Money, War and Computers} (B. G. Tuebner,
Stuttgart, Leipzig, 1999).

\bibitem{IA} S. N. Dorogovtsev, J. F. F. Mendes and A. N.
Samukhin, Phys. Rev. Lett. \textbf{85}, 4633 (2000).

\bibitem{Assort} A. Barrat and R. Pastor-Satorras, Phys. Rev. E
\textbf{71}, 036127 (2005).





\bibitem{DM} S. N. Dorogovtsev and J. F. F. Mendes, Europhys. Lett.
\textbf{52}, 33 (2000).

\bibitem{MS1} W.-X. Wang, B. Hu, T. Zhou, B.-H. Wang, and Y.-B.
Xie, Phys. Rev. E \textbf{72}, 046140 (2005).

\bibitem{aging} S. N. Dorogovtsev and J. F. F. Mendes, Phys. Rev. E
\textbf{62}, 1842 (2000).



\end{thebibliography}
\end{document}